\documentstyle{acmconf}

\newcommand{\mfig}[1]{Figure~\ref{#1}}

\newcommand{\mthm}[1]{Theorem~\ref{#1}}

\newcommand{\msct}[1]{Section~\ref{#1}}
\newcommand{\blst}{\begin{list}{}{}}
\newcommand{\elst}{\end{list}}
\newcommand{\ben}{\begin{enumerate}}
\newcommand{\een}{\end{enumerate}}
\newcommand{\bit}{\begin{itemize}}
\newcommand{\eit}{\end{itemize}}
\newcommand{\bl}{\item}

\newcommand{\mb}[1]{{\bf #1}}

\newcommand{\ms}[1]{{\sf #1}}

\newcommand{\textit}[1]{{\it #1}}
\newcommand{\join}{\bowtie}

\newcommand{\eat}[1]{}

\newcommand{\alg}{{Algorithm}}
\newcommand{\alga}{{\alg}~A}
\newcommand{\algb}{{\alg}~B}
\newcommand{\algc}{{\alg}~C}
\newcommand{\algd}{{\alg}~D}

\newcommand{\emph}[1]{{\em #1\/}}

\newcommand{\floor}[1]{\lfloor {#1} \rfloor}
\newcommand{\bfloor}[1]{\left\lfloor {#1} \right\rfloor}

\newcommand{\eg}{e.g.,~}
\newcommand{\ie}{i.e.,~}

\newcommand{\nb}{{b}} 

\newcommand{\nbg}[1]{{b_{#1}}}

\newcommand{\nl}{\ms{NL}}
\newcommand{\sm}{\ms{SM}}
\newcommand{\pl}{\ms{P}}
\newcommand{\gv}{ : }

\def\bbbu{{\mathchoice {\setbox0=\hbox{$\displaystyle\rm U$}\hbox{\hbox
to0pt{\kern0.4\wd0\vrule height1\ht0\hss}\box0}}
{\setbox0=\hbox{$\textstyle\rm
U$}\hbox{\hbox to0pt{\kern0.4\wd0\vrule height1\ht0\hss}\box0}}
{\setbox0=\hbox{$\scriptstyle\rm U$}\hbox{\hbox to0pt{\kern0.4\wd0\vrule
height1\ht0\hss}\box0}} {\setbox0=\hbox{$\scriptscriptstyle\rm
U$}\hbox{\hbox
to0pt{\kern0.4\wd0\vrule height1\ht0\hss}\box0}}}}

\newcommand{\ds}{\displaystyle}

\newcommand{\pct}{\%\ }

\newtheorem{THEOREM}{Theorem}[section]
\newenvironment{theorem}{\begin{THEOREM} \hspace{-.85em} {\bf :} }%
                        {\end{THEOREM}}
\newtheorem{LEMMA}[THEOREM]{Lemma}
\newenvironment{lemma}{\begin{LEMMA} \hspace{-.85em} {\bf :} }%
                      {\end{LEMMA}}
\newtheorem{COROLLARY}[THEOREM]{Corollary}
\newenvironment{corollary}{\begin{COROLLARY} \hspace{-.85em} {\bf :} }%
                          {\end{COROLLARY}}
\newtheorem{PROPOSITION}[THEOREM]{Proposition}
\newenvironment{proposition}{\begin{PROPOSITION} \hspace{-.85em} {\bf :} }%
                            {\end{PROPOSITION}}
\newtheorem{DEFINITION}[THEOREM]{Definition}
\newenvironment{definition}{\begin{DEFINITION} \hspace{-.85em} {\bf :} \rm}%
                            {\end{DEFINITION}}
\newtheorem{CLAIM}[THEOREM]{Claim}
                            {\end{CLAIM}}
\newtheorem{EXAMPLE}[THEOREM]{Example}
\newenvironment{example}{\begin{EXAMPLE} \hspace{-.85em} {\bf :} \rm}%
                            {\end{EXAMPLE}}
\newtheorem{REMARK}[THEOREM]{Remark}
\newenvironment{remark}{\begin{REMARK} \hspace{-.85em} {\bf :} \rm}%
                            {\end{REMARK}}

\newcommand{\thm}{\begin{theorem}}
\newcommand{\lem}{\begin{lemma}}
\newcommand{\pro}{\begin{proposition}}
\newcommand{\dfn}{\begin{definition}}
\newcommand{\rem}{\begin{remark}}
\newcommand{\xam}{\begin{example}}
\newcommand{\cor}{\begin{corollary}}
\newcommand{\prf}{\noindent{\bf Proof:} }
\newcommand{\ethm}{\end{theorem}}
\newcommand{\elem}{\end{lemma}}
\newcommand{\epro}{\end{proposition}}
\newcommand{\edfn}{\bbox\end{definition}}
\newcommand{\erem}{\bbox\end{remark}}
\newcommand{\exam}{\bbox\end{example}}
\newcommand{\ecor}{\end{corollary}}
\newcommand{\eprf}{\bbox\vspace{\topsep}}
\newcommand{\bbox}{\vrule height7pt width4pt depth1pt}

\def\bbbc{{\mathchoice {\setbox0=\hbox{$\displaystyle\rm C$}\hbox{\hbox
to0pt{\kern0.4\wd0\vrule height0.95\ht0\hss}\box0}}
{\setbox0=\hbox{$\textstyle\rm C$}\hbox{\hbox
to0pt{\kern0.4\wd0\vrule height0.95\ht0\hss}\box0}}
{\setbox0=\hbox{$\scriptstyle\rm C$}\hbox{\hbox
to0pt{\kern0.4\wd0\vrule height0.95\ht0\hss}\box0}}
{\setbox0=\hbox{$\scriptscriptstyle\rm C$}\hbox{\hbox
to0pt{\kern0.4\wd0\vrule height0.95\ht0\hss}\box0}}}}
\newcommand{\cst}{{\bbbc}}
\newcommand{\env}{{\bf v}}
\newcommand{\V}{{\cal V}}

\newcommand{\ec}{\mb{E}_{\cst}}

\newcommand{\E}{\mb{E}}

\newcommand{\commentout}[1]{}
\begin{document}

\title{Least Expected Cost Query Optimization:\\ An Exercise in 
Utility}
\author{
Francis Chu%
\thanks{Work supported in part by NSF under
grant IRI-96-25901.}\ \ \ \ 
Joseph Y. Halpern%
\footnotemark[1]
\ \ \ \ Praveen Seshadri\thanks{Work supported in part by NSF under
grant IRI-96-?????.} \\
Department of Computer Science\\
Upson Hall, Cornell University\\
Ithaca, NY 14853-7501, USA
\\[.125in]
{\tt \{fcc,halpern,praveen\}@cs.cornell.edu}
}
\maketitle

\begin{abstract}
We identify 
two unreasonable, though standard, assumptions made by database query
optimizers that can adversely affect the quality of the chosen evaluation
plans.
One assumption is that it is enough to optimize for the \emph{expected}
case---that is, the case where various
parameters
(like available memory)
take on their expected value.
The other assumption is that the parameters are constant throughout the
execution of the query.
We present an algorithm based on the
``System~R''-style query optimization algorithm that 
does not rely on these assumptions.  
The algorithm we present chooses the plan of the \emph{least expected cost}
instead of the plan of least cost given some fixed value of the parameters.
In execution environments that exhibit a high degree of
variability,
our techniques
should result in 
better performance.
\thispagestyle{empty}
\end{abstract}

\section{Introduction}
\label{s:intro}
A database query is specified declaratively, not procedurally.  Given a
query, it is the job of the DBMS to choose an appropriate evaluation
plan for it.  This task is performed by a cost-based query
optimizer.  In theory, the task of the optimizer is 
simple: 
it performs a search among a large space of equivalent plans for the
query,
estimates the cost of each 
plan,
and returns
the plan of least cost.

In practice, things are not so simple. There are two major
problems: (1) there are far too many possible plans for an optimizer to
consider
them all,
and (2) accurate cost estimation is virtually
impossible, since it requires detailed \emph{a priori} knowledge of the
nature
of the data and the run-time environment.
Because query optimization is such a critical component of
a database system---queries are typically optimized once and then evaluated
repeatedly, often
over many months or years---much effort has gone into dealing well with
these
problems. The query optimizer has become one of the most complex software
modules
in a DBMS.


Dynamic programming techniques are typically used to deal
with the first problem~\cite{Selinger:1}, although randomized algorithms
have also been proposed~\cite{swami:sigmod89,Ioannidis90}.  As we shall see,
they
apply in
our
approach too, so we focus here on the second problem.
To accurately estimate the cost of executing a particular plan, we need
to estimate the values of various parameters.
Typically, these parameters can be divided into three categories:
\begin{enumerate}
\item Parameters representing properties of the data (cardinalities of
tables, distributions of values, etc.).
The DBMS typically maintains estimates of these parameters.
\item Parameters representing properties of the query components
(\eg sizes of groups, selectivities of predicates). Much research has focused
on 
how
to make accurate estimates of selectivities and result
sizes. These techniques typically use histograms~\cite{Poosala96} (part of the
data properties) or sampling~\cite{sampling:lnss}.
\item Parameters representing properties of the run-time environment
(\eg amount of available memory,
processor speed, multiprogramming level, access characteristics of secondary
storage). These are gathered from observations of the realistic deployment
environments.
\end{enumerate}
If the value of a parameter cannot be exactly predicted
(which is almost always the case, 
especially if it can change even
during the execution of a query),
it can at best be modeled by a distribution.
Current optimizers simply approximate each distribution by using the
mean or modal
value.  They then choose the plan that is cheapest under the assumption
that the parameters actually take these specific values~\cite{Selinger:1}
and remain constant during execution.
We call this the {\em least specific cost (LSC)\/} plan.
\eat{We note that there have been research proposals that suggest
dynamic~\cite{graefe:dynamic}
or parametric~\cite{Yannis:parametric} optimization techniques to address
the
issue of
variations in parameter values. We contrast our work with these prior
approaches in Section~\ref{sec:relatedwork}.}

We propose a very different approach in this paper.
We view the query optimizer as an agent trying to make a decision;
it must choose among different plans.  The standard decision-theoretic
approach is to choose a plan that maximizes expected
utility~\cite{res}.
Here 
utility is essentially the negation of the cost: the lower the
cost, the more attractive the plan.
Thus, we argue that rather than choosing the LSC plan,
query optimization algorithms should be directed towards finding the
plan of {\em least expected cost (LEC)}.


\subsection{A Motivating Example}
The following example should help motivate the use of LEC plans.
For pedagogical reasons, we focus in this example, and throughout 
most of the
paper, on only one parameter, the amount of available
memory, which is known to be difficult to predict accurately
in practice~\cite{lohman:pers98}.
Thus we assume that everything else (\eg predicate selectivities) is known.
We 
consider the case of more than one parameter in \msct{sec:multi}.
\xam
\label{xam1}
Consider a query that requires a join between tables $A$
and $B$, and the result needs to be ordered by the join column.  $A$ has
1,000,000
pages, $B$ has 400,000 
pages,
and the result has 3000 pages.
Consider the following two evaluation plans:

\begin{itemize}
\item \emph{Plan 1}: Apply a sort-merge join to $A$ and $B$. The optimizer
cost formulas tell us that if the available buffer size is greater than 1000
pages (the square root of the larger relation), the join requires two
passes over
the relations~\cite{shapiro:joins}. If there are fewer than 1000 pages
available,
it requires
at least another pass. Each pass requires that
1,400,000
pages be
read
and written.
\item \emph{Plan 2}: Apply a Grace hash-join~\cite{shapiro:joins} to $A$ and
$B$
and then sort
their result. We know that if the available buffer size is
greater than 633
pages (the square root of the smaller relation), the hash join requires two
passes over
the input relations. The subsequent sort also incurs additional overhead,
especially if the data does not fit in the buffer.
\end{itemize}
If the available buffer memory is accurately known, it is easy to
choose between the two plans (Plan 1 when more than 1000 pages
are
available and Plan 2 otherwise).
However, assume that the available memory is estimated to be 2000 pages
80\pct of the
time and 700 pages 20\pct of the time.  (This distribution is 
obtained by observing the actual query execution environment.)
Current optimizers assume one specific memory value (in this case, 2000
pages as a
modal value, or 1740 pages as a mean value). In either case, the plan chosen
would be Plan 1. However, we claim that Plan 2 is likely to be cheaper on
average across a large number of evaluations. The intuition is 
simple: In 80\pct
of the runs, Plan 2 is slightly more expensive than Plan 1 (the extra
expense
arises in sorting the small result), whereas in 20\pct of the cases, Plan 1
is
far
more expensive than Plan 2 (the extra expense comes from an extra
pass over the data). 
On average, we would expect Plan 2 to be preferable.
\exam

Example~\ref{xam1} shows the flaw that arises if parameter distributions
are characterized by a single expected 
value:
The least cost plan
found based on
this value is not necessarily the plan of least expected cost. Indeed,
whenever there are discontinuities in cost formulas (as is the case
with
database join algorithms), such an effect is likely to arise.

\subsection{Contributions}
The contributions of this paper are:
\begin{enumerate}
\item
We introduce LEC plans, which are
guaranteed to be at least as good as (and
typically better than) any specific LSC plan.
\eat{That is,
if the distribution we use is an accurate model of the distribution of
the parameters that is encountered in practice and our dynamic
programming
algorithm indeed produces the best plan for each specific setting of the
parameters, than the LEC plan will be at least as good.
}
\eat{The greater
the run-time variation in the values of parameters that
affect the cost of the query plan, the greater the cost advantage of
the LEC plan is likely to be.
}
\item
We show how LEC query optimization can take into account parameters that
have a constant value during any specific query execution
(\emph{static parameters}), and also those that vary during execution
(\emph{dynamic parameters}).
Moreover, it can be applied either at compile-time or at
start-up time.
\item
We show how traditional dynamic programming query
optimizers can be easily extended 
to produce the LEC plan.
The extension increases the cost of query optimization by a factor
depending on the granularity of the parameter distribution.
\eat{
overhead over the cost of
computing the LSC plan, and can be implemented
in existing systems.
}
\item We extend LEC query optimization to handle queries where the 
selectivity of each predicate is a parameter modeled by a distribution.
\end{enumerate}

Depending on the actual distribution of parameters that arises in
practice, the LEC plan can
be
much more efficient on average than any specific LSC plan.
The greater the run-time variation in the values of parameters that affect
the
cost of the query plan, the greater the cost advantage of the LEC plan is
likely to be.  If such parameter distributions are common, it should be well
worth implementing this approach in commercial database systems.


The rest of this paper is organized as follows.  In the next section, we
review
the traditional approaches to query optimization and discuss related work.
In
Section~\ref{s:util},
we introduce LEC query optimization and
discuss how LEC plans may be generated 
in practice by a typical DBMS.
To this end, we consider extensions to the widely used System~R
query optimization algorithm.
Our algorithms apply both in the case where parameters are static and
(under some simplifying assumptions) the case where they are
dynamic.
%
In Section~\ref{s:con} we discuss the simplifying assumptions we make in
the paper and possible future directions.

\section{Background}
\label{s:bgd}
\subsection{Standard Query Optimization}
There are three basic approaches proposed for query optimization algorithms:
\begin{itemize}
\item \emph{Bottom-Up Optimization}: This is a synthetic approach in
which a suitable plan is created by starting from the stored tables and
building increasingly larger plans until a plan for the entire query is
formed.
\item \emph{Top-Down Optimization}: This is a divide-and-conquer approach
in which the entire query is 
divided
into pieces, each piece is optimized, and
then the pieces are put together to form the query plan.
\item \emph{Transformational Optimization}: This starts with some valid
complete query plan, and repeatedly transforms it into a different 
valid complete plan. At every stage, the plan of least cost is retained.
\end{itemize}
Every query optimizer uses some element of each approach. Typically, a query is
divided into a graph of ``query blocks'' and some transformational
optimizations are performed 
on
the query blocks~\cite{Pirahesh:1}. Each
query block in the graph is then typically optimized almost independently.  A
specific kind of query block, the
\emph{SELECT-PROJECT-JOIN} or \emph{SPJ}
block, has received a lot of attention, because it occurs in many queries and
involves expensive join operations. The optimization of an SPJ block itself
could use any of the three basic approaches. Most commercial database systems
use a bottom-up optimizer based on dynamic programming.
This approach was first suggested in the System~R project~\cite{Selinger:1}.
We
now briefly describe how it works.  In later sections of the paper, we
describe
variations of this algorithm that find LEC plans.

\subsection{The System~R Approach}
Suppose we are given $n$
relations, $A_1, \ldots, A_n$,
whose join we want to compute.
For ease of exposition, assume that there are join predicates between every
pair of relations.  (This is not very realistic, but one can always assume
the
existence of a trivially true predicate.)  Three basic observations
influence
the algorithm:
\begin{enumerate}
\item Joins are commutative.
\item Joins are associative.
\item The result of a join does not depend on the algorithm used to
compute it. Consequently, dynamic programming techniques may be
applied.\footnote{It is well-known that this is not quite accurate; the
physical properties (``interesting orders'') of the join result depend
on the specific plan used to create it. This requires simple extensions
of the optimization algorithm, as described in~\cite{Selinger:1}.
We ignore this issue here, since our solutions
apply without change in the presence of these extensions.}
\end{enumerate}


The System~R optimizer also applies some heuristics that further limit
the space of plans considered.
Of particular relevance to this paper
are the following two heuristics:
\begin{enumerate}
\item Only binary join algorithms are considered. Consequently, a
three-relation join evaluation plan involves the combination (\ie join) of a
two-relation join result and a stored relation.
\item To find the best plan for a $k$-relation join, the only
plans considered are those that first involve joining some subset of
$k-1$ of these relations and then adding in the $k$th.
Other possible approaches (for example, considering the best plan for
joining a subset of $k-2$ of the relations, joining the remaining two
relations, and then joining the results)
are not considered.
\end{enumerate}
Given these two heuristics, System~R is essentially trying to find the
permutation $\pi$ of $\{1, \ldots, n\}$ that produces the best plan
of the form
$$
(( \cdots ((A_{\pi(1)} \join A_{\pi(2)}) \join A_{\pi(3)}) \cdots )
\join A_{\pi(n)}).
$$
Such plans are called {\em left-deep\/} plans.

Conceptually, we can think of the System~R optimizer
as working on a dag
with a single root
(node of indegree 0).
Each node in the dag is labeled by a subset $S$ of $\{1, \ldots, n\}$.  The
label of the 
root
is the empty set. The nodes at depth
$k$ are labeled by the subsets of $\{1, \ldots, n\}$ of cardinality $k$.
There is an edge from a node at depth $k-1$ labeled by $S'$ to a node at
depth $k$ labeled by $S$ iff 
$S' \subseteq S$.
Fix a setting of the parameters.
Associated with
the node labeled $S$ is the best
left-deep
plan to
compute the {\em join over $S$\/} (\ie
$\join_{i\in S} A_i$, the join of the relations with indices in $S$)
for that setting of the parameters.
This plan is determined inductively as follows.
Initially, the algorithm determines the best plan to access each of the
individual relations.  In the next step, the algorithm examines all
possible
joins of two relations. For each pair of relations, several different
join
algorithms are considered and the cheapest evaluation plan is retained.
Assume inductively that we have associated with each node up to depth
$k$ the plan for computing the join associated with that node.
To compute the best plan for a node $S$ at depth $k+1$, consider each $j 
\in S$ and let $S_j = S - \{j\}$ and let $B_j = \join_{i \in S_j} A_i$.
For each $j$, let $C_j$ be the sum of the cost of the 
best plan for accessing $A_j$, the cost of the best plan for computing
$B_j$ (which we have already determined), and the cost of the best plan
for computing $B_j \join A_j$.  We then find the $j$ for which $C_j$ is
minimal.  This gives us the the best plan for computing
$\join_{i \in S} A_i$ as well as its cost.
Note that at phase $(k+1)$, 
only results from phase $k$ are utilized.  
At phase $n$, we label the root by the best plan to compute the join
(and its cost).  
Although this approach takes time and space
exponential in $n$, $n$ is usually small enough in practice to make this
approach feasible.

To summarize, we 
have:
\thm\label{Ropt} The System~R optimizer computes the 
LSC 
left-deep plan for a specific setting of the parameters. \ethm
%
(We remark that a formal proof of Theorem~\ref{Ropt} can be provided
along the lines of the proof of Theorem~\ref{LECopt} given below.)

\subsection{Previous Work on Dealing with Uncertainty of Parameter Values}
\label{sec:relatedwork}
It is widely recognized that query optimizers often make poor decisions
because
their compile-time cost models use inaccurate estimates of various
parameters. There have been several efforts in the past to address this
issue. We categorize them as (a) strategies that make decisions at the start
of
query execution
and
(b) strategies that make decisions during query execution.

Let us assume that there are some parameters that cannot be predicted
accurately at compile-time, but that are accurately known when a query
begins execution. An example of such a parameter is the number of
concurrent users. 
Let us further assume 
that the value of this parameter
stays constant during the execution of the query. In this case, we are
aware of three kinds of strategies:
\begin{itemize}
\item A trivial strategy is to perform query optimization just before
query execution. This is the approach used in database systems like
Illustra~\cite{illustra:usermanual}, but is not particularly efficient,
since
the query may be executed repeatedly.
\item Another strategy is to find the best execution plan for every possible
run-time value of the parameter. This requires much additional work at
compile-time, but very little work at query execution time (a simple table
lookup to find the best plan for the current parameter value).
In~\cite{Yannis:parametric}, the authors suggest using randomized
optimization to reduce the compile-time optimization effort.
\item \cite{graefe:dynamic} suggests a
hybrid strategy that performs some of the search activity at compile-time.
Any
decisions that are affected by the value of the parameter are deferred to
start-up time through the use of ``choice nodes'' in the query evaluation
plan.
\end{itemize}

For parameters that cannot be accurately predicted at the start of
query execution (like predicate selectivities), these strategies are
clearly inapplicable. We are aware of 
four
other strategies that address
this case; all involve a 
potential
modification of query execution.
\begin{itemize}
\item \cite{Kabra98} proposes using a regular query optimizer to generate a
single plan, annotated with the expected cost and size statistics at all
stages
of the plan. These statistics are affected by the choice of parameter value.
During actual query execution, the expected statistics are compared with the
measured statistics. If there is a significant difference, the query
execution
is suspended and re-optimization is performed using the more accurate
measured
value of the parameter.
\item \cite{rdb:dyn} implements an interesting variant of this idea. In
order
to execute a query, multiple query plans are run in 
parallel.
When one plan finishes or makes significant progress, the other competing
plans are killed. This strategy assumes that resources are plentiful (and
so can be wasted), and is applied only to subcomponents of the query
(typically to individual table accesses).
\item \cite{ufa98} examines a very specific kind of parameter
variation: the cost of accessing a table across an occasionally faulty
network, such as the Internet.
Their strategy 
reoptimizes the query when the system recognizes during
query execution that the source of 
a
table is not
available.
Instead of restarting the query like~\cite{Kabra98}, the remainder of the query
plan is adjusted (``scrambled'') to try to make forward progress.
\item \cite{sbm93} focuses on 
uncertainties
that can be 
reduced
by sampling 
(more specifically, the
uncertainty 
of the selectivity of a 
predicate).  They use decision-theoretic 
methods
to pre-compute scenarios where
it may be 
worthwhile
to do sampling
(since sampling itself comes with a cost).
If such a scenario arises, they do the
sampling and modify the plan as appropriate, given the result of the sampling.
We remark that this approach is the one perhaps closest to 
that 
advocated here
in its view of query optimization as a decision problem and its aim of
minimizing expected cost.
The techniques of \cite{sbm93} can be combined with those suggested here.
\end{itemize}

\eat{
PhD thesis & journal article on using expectation for query optimization:

 Kevin Seppi : UT Austin (Mech Engg) w/ Wesley Barnes, Carl Morris
    TR#: QRP 89-19

 ORSA Journal on Computing (1993) pp 410-419

 Bayesian Networks for QOpt
}

Note that these approaches all involve making some decision after compile-time.
The way they deal with uncertainty is to wait until they have more information.
We deal with uncertainty by treating the parameters as random variables, so our
approach can be applied completely at compile-time, as well as at start-up time
or run-time. When our approach is applied at compile-time, the size of the
query plan created does not increase as with some of these approaches.

\section{LEC Query Optimization}
\label{s:util}

\subsection{The Formal Model}
\eat{
LEC query optimization is based on the observation that
whenever queries are optimized, predictions must
be made about parameter values. These predictions
typically
take the form
of probability distributions.}


%

Fix a query.
Like~\cite{Yannis:parametric}, we start by assuming that there is a cost
function $\cst$ that takes two arguments, a plan $p$ and a vector $\env$ of
values of relevant parameters, and returns a cost.  Intuitively,
$\cst(\pl,\env)$ is the cost of executing plan $p$ under the assumption that
the relevant variables take the values $\env$.  The standard (LSC) approach is
to choose a fixed value of $\env$---usually the expected value of the
parameters---and find a $p$ with the least cost.
We assume instead that
there is
a probability measure on the space $\V$
of possible values of the parameters.
\eat{
Note that it is not unreasonable to make this assumption about parameters
representing properties of the data (such as cardinality), since the DBMS
already keeps statistics about these parameters.  The same is true for
parameters pertaining to the run-time environment, since these are obtained via
observations of the environment.  This leaves parameters about properties of
the query components (\eg selectivity of predicates, size of groups, etc.).
One way to deal with this is to keep statistics across different executions.
Once we
}
Given $\cst$ and $\Pr$, we can compute the expected cost for each plan $p$.
Let $\ec(\pl)$ denote the expected cost of $p$ (with respect to cost
function $\cst$ and probability $\Pr$); as usual, we have
$$\ec(\pl) = \sum_{\env \in \V} \cst(\pl,\env)\Pr(\env).$$
%
Following standard decision-theoretic approach, our goal is to find
the {\em LEC plan\/}, the one whose expected cost is least.
If the distribution
$\Pr$
is an accurate model of
the distribution of the parameters that is encountered at run-time, and
the cost estimates $\cst$ are accurate then, by definition, the
expected
execution cost of
the LEC plan is at least as low as that of any specific LSC plan.

The
goal of finding the LEC plan
makes sense both at compile-time and at start-up time.  At start-up time, 
the distribution of parameters will typically be different from the one at
compile-time.
For some parameters, the distribution may be more concentrated
around one value.  However, it is unlikely that there will be complete
information about all the values of the relevant parameters, even at
start-up
time.  This is particularly true about parameters whose values may change as
we
execute the query.  Note that the LEC approach applies 
to such
dynamic parameters
as well.  
We need to use a probability distribution over all
possible sequences of parameter values during the execution and then again
compute the plan with least expected cost.
We explore the details of this approach 
in Section~\ref{sec:variation}. 
Until then,  we assume 
that parameters do not change value during query execution.

\eat{
Note that if the distribution we use is an accurate model of the
distribution of the parameters that is encountered in practice and our
cost estimates are accurate, then the LEC plan is guaranteed to be at
least as good as the LSC plan, for any specific setting of the parameters.
}

This leads to some obvious questions:
\begin{enumerate}
\item How do we get the probability distributions?
\item How do we get the cost estimates?
\item How do we compute the LEC plan?
\end{enumerate}
With regard to the first question, as we noted in the introduction, the DBMS
in
practice is constantly gathering statistical information.  We believe that
the
statistics can be enhanced to provide reasonable estimates of the relevant
probabilities, although this is certainly an area for further research.  For
the purposes of this paper, we assume without further comment that the
probabilities are available.  
As for the second question, we 
are making the
same assumptions about costs that are made by all standard query optimizers.
Finally, with regard to the last question, in the remainder of this section,
we
provide a number of ways to modify the standard optimizers so as to produce
the
LEC plan (or a reasonable approximation to it).  These approaches vary in
their
need to modify the underlying query optimizer, the quality of the plan
produced, and the underlying assumptions about the distribution.

\subsection{\alga: Using a Standard Query Optimizer as a Black Box}
\label{sec:lec:blackbox}
For ease of exposition,
we assume 
in the next few sections
that the only relevant parameter is the
amount of available memory, so we take $\env$ to represent this
quantity.
This assumption is dropped in Section~\ref{sec:multi}
We assume that we can partition the 
distribution of the amount of available memory into a small number (say $\nb$)
of buckets such that the cost of a plan is likely to remain
relatively
constant within a bucket.
For example, in Example~\ref{xam1}, the appropriate buckets are the
intervals 
$[0,633)$, $[633,1000)$, and $[1000,\infty)$.
\eat{While we believe that in practice it is often the case that such
buckets
are relatively easily defined, we may well want to extend our algorithm
so that it learns a good subdivision, using the cost function; we
discuss this issue in more detail in the full paper.
}
Choosing the buckets appropriately can be nontrivial;
we discuss this issue in more detail in Section~\ref{sec:tuning}.  
We can identify the standard approach to doing query optimization with the
special case where there is only one bucket.  Once we have chosen the buckets,
we pick a representative from each bucket; call them $m_1, \ldots, m_\nb$.
Finally, we assume that we have a probability measure $\Pr$ such that
$\Pr(m_i)$ characterizes how likely we are to run the query
in the
$i$th
bucket.

Given these assumptions, there is a straightforward approach to finding good
approximations to the LEC plan that uses a standard query optimizer as a black
box.  Suppose we want to compute $A_1 \join \cdots \join A_n$.  Assume that
memory stays constant during the execution of the plan.  

\blst
\item[{\bf \alga}]
\verb+  +

For each value $m_i$
of the memory parameter, we run the optimizer under the assumption that $m_i$
is
the actual
amount of
memory available.  This gives us $\nb$ 
candidate plans.  
We then compute the expected cost of each candidate, and choose the one with
least expected cost.
\elst

As long as the expected value of memory used by the traditional LSC approach is
one of the $\nb$ possible values we consider (and, without loss of generality,
we can assume it is), then we are guaranteed to end up with a plan whose
expected cost is no higher than that of the plan chosen by the traditional
approach.
We
assume that, in practice, the actual LEC plan will have a cost close to the
optimal plan for some value $m_i$ of memory.  To the extent that this is true,
\alga{}
gives us a good approximation to the LEC plan.

The cost of 
\alga{}
is the cost of $\nb$ invocations of the optimizer, plus the cost of evaluating
the expected cost of each candidate plan. Each candidate plan has $n-1$ joins,
and each has to be costed for $\nb$ different memory sizes to determine the
expected cost. There are $\nb$ such candidates, leading to a total cost of
$O((n-1) \nb^2)$, which should be much smaller than the cost of candidate
generation, $O(\nb n 2^{n-1})$.  
Consequently, the approximate cost of 
\alga{}
is $\nb$ times the cost of a single optimizer invocation.

Note that it makes sense to use 
\alga{}
at start-up time as well as at compile-time; we simply use the appropriate
distribution over memory sizes when checking to see which candidate plan is
best.  We can also combine these ideas with the parametric query optimization
approach of~\cite{Yannis:parametric}.  We can precompute the best expected plan
under a number of possible distributions (ones that give good coverage of what
we expect to encounter at run-time), and store these expected plans, for use at
query execution time.

While this approach has the advantage of not requiring any change to the
optimizer whatsoever,
it has two major drawbacks.  The first is that it requires us to
prespecify the buckets (this point should become clearer in
Section~\ref{sec:tuning}); the second is that it 
may
not actually
return
the LEC plan.
It is conceivable that a plan not optimal for any $m_i$ actually does
better on average 
than
any candidate 
considered
by the algorithm above.
For example, the plan that is second-best for some memory size may do better
on
other memory sizes than the best plan for that memory size, and so may do
better in expectation.  We now present a simple modification of this
approach
that generates more candidate plans, although it has the disadvantage of
requiring us to modify the basic query optimizer.


\subsection{\algb: Generating More Candidates}
\label{sec:lec:cand}

Suppose that rather than generating the best plan for each memory size
$m_i$,
we generate the top $c$ plans, for some $c > 1$.  It is relatively
straightforward to modify existing query optimizers to do this.  For
concreteness, we show how this can be done with
System~R.
\eat{System~R
outputs the plan with the least cost at each stage of the optimization
process.  We need to modify it so that at each stage it keeps the top $c$
plans for $c > 1$.  We do this recursively.
 For each $A_j$, the optimizer finds the top
$c$ ways to access
$A_j$. When the optimizer wants to compute $A_1 \join \cdots \join
A_{j}$, it will have the top $c$ plans for doing $A_1 \join \cdots \join
A_{j-1}$ and the top $c$ access paths for accessing $A_j$.  It then uses
these
to compute the top $c$ plans for $A_1 \join \cdots \join A_j$.  So in
the end,
we have the top $c$ plans for $A_1 \join \cdots \join A_n$.  Do this
for each
$m_i$.  This gives us $c k$ candidates.

Suppose we go through the $m_i$s in increasing order.  Then when we
compute the $c$ best plan for $A_1 \join \cdots \join A_j$, for $m_i$ we
have
already done the computation for $m_1, \ldots, m_{i-1}$ and we can use these
for the computation of $m_i$.  If the steps are done in parallel, we can
take
combinations of plans from $m_{j_1}$ and $m_{j_2}$ such that
 $m_{j_1} + m_{j_2} \leq m_i$; if the steps are done in serial, then we just
need $m_{j_1}, m_{j_2} < m_i$.
}

\blst
\item[{\bf \algb}]
\verb+  +

Assume inductively that we have associated with each node up to depth $k$ in
the dag the top $c$ plans for computing the join associated with that node.  To
compute the top $c$ plans for a node $S$ at depth $k+1$, consider each $j \in
S$ and let $S_j = S - \{j\}$, as before.  We consider the top $c$ plans for
computing the join over $S_j$ and the top $c$ plans for accessing $A_j$;
combining them
using each possible join method gives us the top $c$ plans for computing the
join over $S$ if we join $A_j$ last.
\elst
While it seems that there are $c^2$
combinations of plans that need to be considered here
 for each join method,
\eat{as we show in the
full paper, it suffices to consider at most $c + c \log{c}$
combinations of plans to produce the top $c$ plans.
}
the actual number of combinations is lower.
\pro
It suffices to consider at most $c + c\log{c}$ combinations of plans 
for each join method
to produce the top $c$ plans.
\epro
\prf
Suppose $s_1, \ldots, s_c$ are the top $c$ plans for computing $S_j$
(sorted in
increasing order of cost) and $a_1, \ldots, a_c$ are the top $c$ plans
for accessing $A_j$ (again sorted in increasing order of cost).  Note
that the cost
of the combination $(s_i, a_k)$ is no higher than the combination
$(s_{i'}, a_{k'})$ if $i \leq i'$ and $k \leq k'$; so there are (at
least) $i k - 1$ combinations with cost no higher than $(s_i, a_k)$.
Thus we only have to consider $(s_i, a_k)$ for $i k \leq c$, since $i k
> c$ implies we can get (at least) $c$ plans 
at least as good as $(s_i, a_k)$. 

Note that $i k \leq c$ implies $i \leq c/k$, so we only need to consider the
top $\floor{c/k}$ entries of the $k$th column if we arrange the combinations
in
a $c \times c$ matrix.  Thus the total number of entries we need to consider
is
\[
\sum_{k=1}^c \bfloor{\frac{c}{k}} \leq \sum_{k=1}^c \frac{c}{k} = c
\sum_{k=1}^c \frac{1}{k}.
\] 
Recall that
\[\sum_{x=1}^c \frac{1}{x} < 1 + \int_{1}^c \frac{1}{x} dx = 1+\log{c}.
\] 
Thus the total number of entries we need to consider is at most $c + c
\log{c}$.
\eprf
\\
To compute the best $c$ plans 
using a particular join method
for joining $S_j$ and $A_j$, we
must first evaluate the cost formula for the join method.
Note that all the $c$ variants of each input have the very same
properties, and so behave identically with respect to the cost formula.
Consequently, the only difference between the $c + c\log{c}$ combinations 
arises from the sum of the costs of the two input plans. Consequently, the
cost of checking these combinations is expected to be small compared to the
cost of evaluating a cost formula.
By considering all $j \in S$, we get $j$ lists of
$c$ top plans.  We then take the top $c$ plans from the combined list.
This extension can be easily implemented and is a relatively small
and localized change to current optimizers.
\eat{
In the full paper, we show that this takes $O(c \log {c})$ times as long
as just generating the top plan.
}
\thm 
\algb{}
computes the top $c$ left-deep plans for each of the
$\nb$ choices of parameter values at $\alpha \nb$ times the cost
of computing the single best left-deep plan for one specific setting of
the parameters, for some small constant $\alpha > 0$. \ethm

Once we have the top $c$ plans for each of the $\nb$ memory sizes, we can
then again compute the expected cost of each of these $c\nb$ plans, and
choose the plan of least expected cost.
As we showed in the previous section, the computation of expected costs
of candidate plans is small compared to the cost of candidate generation.
In this case, the number of candidates is increased by a factor of $c$,
but we still expect 
\algb{}
to be roughly $\nb$ times as expensive
as a single optimizer invocation.

While 
\algb{}
generates more candidates (and thus is more likely to end up with a good
approximation to the LEC plan), it still does not necessarily end up with the
LEC plan.  As we now show, if we are willing to modify the basic query
optimization algorithm further, we can produce the actual LEC plan.

\subsection{\algc: A Generic Algorithm for Computing the LEC Plan}
\label{generic}

We now provide a generic modification of the basic System~R query optimizer
that can directly compute the LEC plan, merging the candidate generation and
costing phases.
We assume inductively that we have associated with each node up to depth
$k$ in the dag the plan with least expected cost for computing the join
associated with that 
node (as well as the expected cost itself).
We further assume that with each node we have
associated a probability distribution over the possible memory
sizes. Intuitively, this is the probability distribution over the available
memory when we reach that node
during an actual execution of a query plan containing the node as a subplan.
If we assume that memory size does not change during the course of executing
the plan, 
and that join operations are not pipelined in the plan,
then the distribution is the same at every node.
\eat{
As we shall see in the next
subsection, we can apply our generic algorithm even if the distribution does
change as we execute the plan (under certain simplifying assumptions).
}
We do away with this restriction in the Section~\ref{sec:variation}.

\blst
\item[{\bf \algc}]
\verb+  +

Again, we proceed inductively down the dag.  
For each stored relation $A_i$, find an LEC access path for it.
To compute the plan with least
expected cost for a node at depth $k+1$ labeled $S$, consider each $j \in S$
and let $S_j = S - \{j\}$.  The expected total cost of $S$ is
the expected cost of computing $S_j$ (which, by assumption, we already have
in
hand) added to the expected cost of joining $S_j$ and
$A_j$,
which we can compute given the probability distribution of available memory.
If we consider a probability distribution over $\nb$ different memory sizes,
this computation requires $\nb$ 
evaluations 
of the cost formula for the join algorithm.
We retain the plan for $S$ with the least expected total cost, discarding
all the other candidates.
\elst

\thm\label{LECopt} 
\algc{}
gives us the LEC  left-deep plan. \ethm
\prf The proof is a straightforward adaptation of the argument for the
correctness of the basic System~R algorithm, using the fact that
expectation distributes over addition.
Suppose $S$ is a nonempty subset of $\{1, \ldots, n\}$.  Let $\pl_S$
denote a left-deep plan for computing $\join_{i \in S} A_i$. 
We can conceptually think of $\pl_S$ as consisting of 
a choice of $j \in
S$, a plan $\pl_S^R$ for accessing $A_j$, and, if $|S| > 1$, a plan
$\pl_S^L$ for computing
$B_j = \ \join_{i \in  S_j} A_i$
and a plan $\pl_S^{\join}$ for computing $B_j \join A_j$.  
If $|S| > 1$ then
\[
\cst(\pl_S,\env) = \cst(\pl_S^L,\env) + \cst(\pl_S^R,\env) +
\cst(\pl_S^{\join},\env).
\]
It follows that 
\[
\ec(\pl_S) = \ec(\pl_S^L) + \ec(\pl_S^R) + \ec(\pl_S^{\join}).
\]
\newcommand{\pa}{\widehat{\pl}}
Let $\pa_S$ be the plan that 
\algc{}
outputs for $S$.  We want to show that $\ec(\pa_S) \leq \ec(\pl_S)$ for all
$\pl_S$. We proceed by induction on $|S|$.

For the base case we have $|S| = 1$.  Suppose $S = \{i\}$.  Then
$\ec(\pa_S) \leq \ec(\pl_S)$, since 
\algc{}
will choose an LEC access path for $A_i$.  Now assume that the claim holds for
all $S$ with $|S| = k$.  Let $S$ be a subset of $\{1, \ldots, n\}$ with $k+1$
elements.  For all $j \in S$, let $\pl(j)$ be a plan such that $\pl(j)^L =
\pa_{S_j}$, $\pl(j)^R = \pa_{\{j\}}$, and $\pl(j)^{\join}$ is an LEC method to
compute the join.  It is clear from the description of 
\algc{}
that $\pa_S \in \{\pl(j) : j \in S\}$ and that $\ec(\pa_S) = \min \{ \pl(j) : j
\in S \}$. Suppose $\pl_S$ is an arbitrary left-deep plan to compute $\join_{i \in
S} A_i$.  Suppose $\pl_S$ computes $\join_{i\in S_j} A_i$ first.  Thus
$\pl_S^L$ computes $\join_{i \in S_j} A_i$.  By the induction hypothesis and
the definition of $\pl(j)$, we see that $\ec(\pl_S^L) \geq
\ec(\pa_{S_j}) = \ec(\pl(j)^L)$.  Furthermore, $\ec(\pl_S^R) \geq
\ec(\pa_{\{j\}}) = \ec(\pl(j)^R)$, since $\pa_{\{j\}}$ is an LEC access
path. Finally, $\ec(\pl_S^{\join}) \geq \ec(\pl(j)^{\join})$, since
$\pl(j)^{\join}$ is an LEC join method.  Thus $\ec(\pl_S) \geq \ec(\pl(j))
\geq \ec(\pa_S)$ as required.
\eprf

If we assume that memory does
not change as we execute the plan, then
the cost of the computation is $\nb$ times the cost of the standard
computation
using a single memory size.  Again, 
\algc{}
works both at compile-time
and
at start-up time.  And again, we can combine these ideas with those of
parametric query optimization, precomputing the LEC plans under various
assumptions about the probability distributions and storing them for use at
start-up time.

\subsection{Dealing with Change During Execution}
\label{sec:variation}
So far we have assumed that the amount of available memory stays constant
throughout the execution of the plan.  This may not be so reasonable if the
query takes a long time
(on
the order of minutes or more).  During the execution,
concurrent new queries may start while old queries may finish.
If we assume that available memory is mainly determined by the number of
queries being run concurrently, then it may well change 
during the execution.
%

To deal with dynamic memory changes, we
assume a probability measure over the possible sequences of memory sizes.  We
then must evaluate each candidate plan with respect to this sequence
and determine its expected cost.
To keep the analysis from becoming too unwieldy, we
%
%
assume that plan execution takes place in phases, each corresponding to
a join in the plan. We assume that memory does not change during the
execution of a phase, but can change between phases. 
If we compute a join over $n$ relations, there are $n-1$ phases.
Thus, we need to consider possible run-time environments corresponding
to each sequence of memory sizes of length $n-1$. We need to use
a probability distribution over all such memory size sequences.

Where does this distribution come from?  Perhaps the most natural
assumption is to assume that we have some distribution over the initial
memory sizes, and that there is a transition probability describing how
likely memory is to change by $m$ units, for each value of $m$.
For simplicity, we assume that this transition probability depends only
on the current memory usage, not on the time. This is a reasonable
assumption for
$24 \times 7$
systems in stable operational mode.

Under these simplifying assumptions, we can apply 
\algc{}
presented in the previous subsection to calculate the LEC plan even with
dynamic memory.  We simply associate the initial distribution with the root of
the dag, and use the transition probabilities to compute the distribution
associated with each node in the tree.
We can then apply the algorithm without 
change, since \algc{} does not rely on the fact that $\env$ consists of a
single memory size.  Note, however, that the complexity of \algc{} is clearly
affected, since the parameter space is potentially much bigger (since if there
are $\nbg{M}$ memory values we consider, there are $\nbg{M}^{n-1}$ sequences
of length $n-1$).

\thm Given the simplifying assumptions above, 
\algc{}
returns the LEC left-deep plan even in the presence of dynamically varying
parameters. \ethm
\prf
While $\env$ now consists of a sequence of memory sizes, the proof of
\mthm{LECopt} works here also, since that proof did not rely on the fact that
$\env$ is a single memory size in any way.
\eprf

\subsection{Dealing with Multiple Parameters}
\label{sec:multi}

We have focused on only one parameter so far (\ie the amount of available
memory).  In
practice, we typically have to deal with a number of parameters.
In this section, we consider the effect of multiple parameters on our
algorithms (specifically, available memory and the selectivities of all the
query predicates). 
Selectivities, in particular, are notoriously uncertain.
We believe that by representing the uncertainty by a probability
distribution and computing the LEC plan, we can ameliorate some of the
difficulty.  Note that the ideas of \cite{sbm93} for deciding when to
sample may also be usefully applied here.
We focus on the static case in this section.  We can apply the idea
from \msct{sec:variation} to deal with the dynamic case.


\begin{figure}
\newcommand{\dashlinestretch}{}
\input{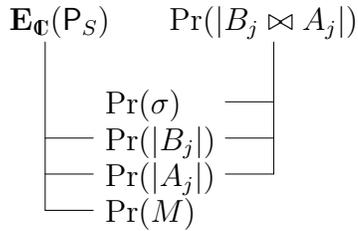}
\begin{center}
\setlength{\unitlength}{0.00083333in}
\begingroup\makeatletter\ifx\SetFigFont\undefined
\def\x#1#2#3#4#5#6#7\relax{\def\x{#1#2#3#4#5#6}}%
\expandafter\x\fmtname xxxxxx\relax \def\y{splain}%
\ifx\x\y   
\gdef\SetFigFont#1#2#3{%
  \ifnum #1<17\tiny\else \ifnum #1<20\small\else
  \ifnum #1<24\normalsize\else \ifnum #1<29\large\else
  \ifnum #1<34\Large\else \ifnum #1<41\LARGE\else
     \huge\fi\fi\fi\fi\fi\fi
  \csname #3\endcsname}%
\else
\gdef\SetFigFont#1#2#3{\begingroup
  \count@#1\relax \ifnum 25<\count@\count@25\fi
  \def\x{\endgroup\@setsize\SetFigFont{#2pt}}%
  \expandafter\x
    \csname \romannumeral\the\count@ pt\expandafter\endcsname
    \csname @\romannumeral\the\count@ pt\endcsname
  \csname #3\endcsname}%
\fi
\fi\endgroup
{\renewcommand{\dashlinestretch}{30}
\begin{picture}(1969,1395)(0,-10)
\path(225,336)(525,336)
\path(225,561)(525,561)
\path(225,111)(525,111)
\path(1350,561)(1650,561)
\path(1350,786)(1650,786)
\path(1350,336)(1650,336)
\path(225,1161)(225,111)
\path(1650,1161)(1650,336)
\put(600,36){\makebox(0,0)[lb]{\smash{{{\SetFigFont{12}{14.4}{rm}$\Pr(M)$}}}}}
\put(600,711){\makebox(0,0)[lb]{\smash{{{\SetFigFont{12}{14.4}{rm}$\Pr(\sigma)$}}}}}
\put(600,486){\makebox(0,0)[lb]{\smash{{{\SetFigFont{12}{14.4}{rm}$\Pr(|B_j|)$}}}}}
\put(600,261){\makebox(0,0)[lb]{\smash{{{\SetFigFont{12}{14.4}{rm}$\Pr(|A_j|)$}}}}}
\put(0,1236){\makebox(0,0)[lb]{\smash{{{\SetFigFont{12}{14.4}{rm}$\ec(\pl_S)$}}}}}
\put(1000,1236){\makebox(0,0)[lb]{\smash{{{\SetFigFont{12}{14.4}{rm}$\Pr(|B_j \join A_j|)$}}}}}
\end{picture}
}
\end{center}
\hrule
\caption{\label{f1}Distributions Needed at Each Node and What Depends on Them}
\end{figure}

When the amount of available memory is the only uncertainty, we need only a
single probability distribution at every node in the dag to allow us to compute
the LEC plan.  When there is more than one uncertain parameter, we may need to
expand that to a joint distribution, whose size may grow exponentially as the
number of parameters.
Since independence often holds in practice or is the default assumption
of existing query optimizers, we assume for the rest of
this section that all parameters of interest are independent.
This simplifying assumption means that we can carry a separate
distribution for each parameter and avoid the exponential blow-up in the
description of the joint distribution.
If are some dependencies between the variables, but not too many,
we can still describe the distribution succinctly using a Bayesian
network \cite{Pearl}.  We believe that the techniques that we present
here will also be applicable to that case.

Note that the standard formulas
for computing the cost of a join plan typically take three parameters:
the sizes of the two relations being joined
and the amount of available memory.  Thus, 
to compute the expected cost of a join method applied to a
particular pair of relations, 
we need just three distributions.
This means that even though we may have many 
more parameters
to deal with, if we apply 
\algc{}
to the multi-parameter setting, at each node only three distributions are
required to compute the LEC plan for that node.
To be able to apply this idea inductively at every node in the dag,
we also need to compute the distribution of the size of the result of the join,
since the parent node of the current node (should there be any) needs that as
the distribution of one of its input size.  In order to compute the 
distribution of the
size of the result, we need to have a distribution of the selectivity of the
join predicates.  Thus at each node, we need 
exactly
four distributions---the three distributions for computing expected cost plus
the distribution for the selectivity of the join predicate---%
no matter how many distributions we start with.
\mfig{f1} shows the distributions we carry at each node and the quantity that
depends on them.

%
%
Here is the generic modified algorithm.  As in \msct{generic}, the algorithm
works on the dag.  

\blst
\item[{\bf \algd}]
\verb+  +

To compute the LEC plan for a node at depth $k+1$ labeled
$S$, consider each $j \in S$ and let $S_j = S - \{j\}$.  Let $B_j = \
\join_{i
\in S_j} A_i$.  We assume inductively that we have the LEC plan for $S_j$ as
well as the distribution of $|B_j|$.  We also assume we have the best way to
access $A_j$ and the distribution of $|A_j|$ after any initial selection.  
We evaluate the cost formula for each 
triple of possible values of $M$, $|B_j|$ and $|A_j|$, and use that to
compute the expected cost of calculating $S_j \join A_j$ for each method.
Taking
$\nbg{X}$ to denote the number of buckets for random variable $X$,
this shows that we need 
$\nbg{M}\nbg{|B_j|}\nbg{|A_j|}$ evaluations to compute
the expected cost of a particular method for computing $S_j \join A_j$.
We can compute the expected cost for each choice of $j$, and retain the
plan of least expected cost.  The total cost of this naive computation
is $\nbg{M}\sum_{j \in S} \nbg{|B_j|}\nbg{|A_j|}$ for each join
algorithm considered.  

To compute the distribution for the size of the result, we 
fix a $j$ and compute, for each triple $(a,b,\sigma)$ of possible values of
$|A_j|$, $|B_j|$, and selectivity $\sigma$ the probability that the join has
size $ab\sigma$.  (Since we are assuming independence, this is just the product
of $\Pr(|A_j| = a)$, $\Pr(|B_j| = b)$, and the probability that the selectivity
is $\sigma$.)  From this we can fill in the distribution for the size of the
result.
\elst
Thus, we require $O(\nbg{|B_j|} \nbg{|A_j|} \nbg{\sigma})$ operations to
compute the distribution.  
Since the size of the result is independent of the choice of $j$, we need to do
this computation for only one $j$; we can choose the $j$ for which
$\nbg{|B_j|}\nbg{|A_j|}$ is minimal.  In practice, we expect that we will have
$\nbg{|A_j|} = \nbg{|B_j|}$ for all $j$, so the choice will not matter.

To summarize, the generic method given above takes
$\sum_{j \in S} \nbg{M} \nbg{|B_j|} \nbg{|A_j|}$ evaluations of the cost
formula and takes $O(\nbg{|B_j|} \nbg{|A_j|} \nbg{\sigma})$ operations to
compute the distribution for $|B_j \join A_j|$.  Can we do better if we have
more knowledge of the cost formula?
Most join methods used in practice have relatively simple cost formulas.  We
pick sort-merge join and page nested-loop join as examples, and demonstrate in
the next two subsections that we can indeed do much better.

\subsubsection{The Case of Sort-Merge Join}

Let $L = \max \{|A|, |B|\}$; the cost of sort-merge join for $A \join B$ is:%
\footnote{Our formulas consider I/O costs only and are based on the analysis
presented in~\cite{shapiro:joins}, simplified to three cases. Commercial
database systems use more complicated formulas, usually represented in the form
of complex code. These are sometimes the result of aiming for too much accuracy
when modeling the algorithm, despite the fact that the parameters used to
instantiate the model are inaccurate. We speculate that a return to simple
formulas in combination with LEC optimization may result in more reliable query
optimizers.}
\[
\cst(\sm,\env) = 
\left\{
\begin{array}{ll}
2\cdot(|A|+|B|)  & \mbox{if } M > \sqrt{L}\\
4\cdot(|A|+|B|)  & \mbox{if } \sqrt[3]{L} < M \leq \sqrt{L}\\
6\cdot(|A|+|B|)  & \mbox{if } M \leq \sqrt[3]{L}\\
\end{array}
\right.
\]
Note that 
\[
\begin{array}{lll}
\ec(\sm)  & \!\! = \!\! & \ec(\sm \gv |A| \leq |B|) \Pr(|A| \leq |B|) + \\
& & \ec(\sm \gv |A| > |B|) \Pr(|A| > |B|).\\
\end{array}
\]  
We now show how the first term can be computed efficiently.  The second term
can be computed analogously; we leave the details to the reader.  Let
$F_b = \E(|A| \gv |A| \leq b) + b$.  Let $Val(X)$ denote the representatives
from the $\nbg{X}$ buckets for variable $X$.  We can rewrite the first summand
as follows:\footnote{We write $X =
x$ as an abbreviation for the statement ``$X$ takes on a value in the bucket
whose representative is $x$''.}
\begin{equation}\label{eceq}
\ds 
\!\!\! 
\sum_{b\in Val(|B|)} \Pr(|B| = b) 
\left(
\begin{array}{ll}
2F_b \Pr(M > \sqrt{b}) & \!\!\!\! +\\
4F_b \Pr(\sqrt[3]{b} < M \leq \sqrt{b}) & \!\!\!\!  +\\
6F_b \Pr(M \leq \sqrt[3]{b}) 
\end{array}
\right).
\end{equation}

We now describe how to compute $\ec(\sm \gv |A| \leq |B|)$ and $\Pr(|A|
\leq |B|)$ in time $O(\nbg{M} + \nbg{|A|} + \nbg{|B|})$.
First we compute $\Pr(M \geq m)$ for each $m \in Val(M)$.  
We can easily compute all $Val(M)$ probabilities in time
$O(\nbg{M})$.
We store these values in a table.  By table lookup, we can then compute $\Pr(M
> b)$, $\Pr(\sqrt[3]{b} < M \leq \sqrt{b})$, and $\Pr(M \leq \sqrt[3]{b})$, for
each value of $b$, in constant time.

Next we compute $\Pr(|A| \leq b)$ for each $b \in Val(|B|)$  
and store these values in a table.  Since $\Pr(|A| \leq b') = \Pr(|A| \leq b) +
\Pr(b < |A| \leq b')$, for $b < b'$, we can compute all of these probabilities
in time $O(\nbg{|A|} + \nbg{|B|})$ (because we need only go through each set of
buckets once).

Then we compute $\E(|A| \gv |A| \leq b)$ for each $b \in Val(|B|)$ and store
these values.  Again, this can be done in time $O(\nbg{|A|} + \nbg{|B|})$,
since $\E(|A| \gv |A| \leq b') = \E(|A| \gv |A| \leq b) \Pr(|A| \leq b) +
\E(|A| \gv b < |A| \leq b') \Pr(b < |A| \leq b')$ for $b < b'$ (and so we only
need to go through each set of buckets once).

We can now compute $\ec(\sm \gv |A| \leq |B|)$ using (\ref{eceq}).  Note we
need only constant time to compute each summand, since for each $b \in
Val(|B|)$ (since $F_b$ can be computed by adding $b$ to $\E(|A| \gv |A| \leq
b)$, which we already have, and it takes only constant time to compute the
probabilities involving $M$, since they have been stored.  Thus, we can compute
the expectation in total time $O(\nbg{M} + \nbg{|A|} + \nbg{|B|})$ (including
the time it take to compute all the values we have stored in the tables).

Finally, we need to compute $\Pr(|A| \leq |B|)$.  This can be done in time
$O(\nbg{|B|})$, since
\[
\Pr(|A| \leq |B|) = \sum_{b \in Val(|B|} \Pr(|A| \leq b) \Pr(|B| = b),
\]
and we have already computed $\Pr(|A| \leq b)$.

The whole computation takes time $O(\nbg{M} + \nbg{|A|} + \nbg{|B|})$,
so the algorithm we informally described is linear in the total number  
of buckets. Note that this algorithm is (asymptotically) optimal, since
we must at least look at the entries in the individual distributions.

We need to carry out this computation for every node in the dag.  (The
$A$ and $B$ above become $B_j$ and $A_j$, using our earlier notation.)  
Note that some of the cost can be amortized over the nodes.  We need  
to do  the computation of $\Pr(M \geq m)$ for $m \in Val(M)$ and 
and $\Pr(A_j \le a)$ for $a \in Val(A_j)$, $j = 1,
\ldots, n$ only once, since these probability distributions do not 
change over the course of the 
execution.
If we precompute these values, then 
the amount of work at each node is only $O(\nbg{|B_j|})$
(for sort-merge).



\subsubsection{The Case of Nested-Loop Join}

As another example, we look at the nested-loop join method.  Let $S = \min
\{|A|, |B|\}$. The cost formula for $A \join B$ using nested-loop join is:
\[
\cst(\nl,\env) = 
\left\{
\begin{array}{ll}
|A|+|B| & \mbox{if } M \geq S+2\\
|A|+|A|\cdot|B| & \mbox{if } M < S+2\\
\end{array}
\right.
\]
As in the previous section, we split $\ec(\nl)$ into two terms: $\ec(\nl \gv
|A| \leq |B|) \Pr(|A| \leq |B|)$ and $\ec(\nl \gv |A| > |B|) \Pr(|A| > |B|)$.
Again we focus on the first term and leave the second term to the reader.  Let
$G_a = \E(|B| \gv a \leq |B|)$, for $a \in Val(|A|)$.  Note that the first term
can be rewritten as follows:
\begin{equation}\label{eceq2}
\ds
\!\!
\sum_{a \in Val(|A|)} \Pr(|A| = a) 
\left( \!\!
\begin{array}{l}
(a+G_a) \Pr(M \geq a + 2) +\\
(a+a G_a) \Pr(M < a + 2)
\end{array}
\right).
\end{equation}
As before, we can do the computation in time $O(\nbg{M} + \nbg{|A|} +
\nbg{|B|})$. 
The procedure is very similar to that for sorted-merge join, so we just sketch
the details here.

We again compute $\Pr(M \geq m)$ for each $m \in Val(M)$.  This takes
$O(\nbg{M})$ steps and enables us to compute $\Pr(M \geq a+2)$ in a constant
number of steps.  Next we compute $\Pr(a \leq |B|)$ for each $a \in Val(|A|)$.
As before, we can compute all $Val(|A|)$ probabilities in time $O(\nbg{|A|} +
\nbg{|B|})$.  Then we compute $\E(|B| \gv a \leq |B|)$ for each $a \in
Val(|A|)$.
Arguments similar to those used in the sort-merge case show that 
this can be done in time
$O(\nbg{|A|} + \nbg{|B|})$. 
We can then compute  $\ec(\nl \gv |A| \leq |B|)$ 
using (\ref{eceq2}).
Again each summand only requires constant time, since we already have $G_a$ and
we can determine the probabilities involving $M$ in a constant number of steps.
Finally, we can compute $\Pr(|A| \leq |B|)$
in time $O(\nbg{|A|})$, just as in the case of sort-merge.

The whole process again takes time $O(\nbg{M}+\nbg{|A|}+\nbg{|B|})$, so the
algorithm is linear in the (total) number of buckets.  

As before, we only considered a single node in the dag.  If we precompute
$\Pr(M \geq m)$ and $\Pr(B \geq b)$, then the amount of work at each node for
nested-loop is $O(\nbg{|A|})$.
\subsubsection{The Distribution of the Result Size}

We showed that the expected cost of specific join methods can be computed in
time linear to the number of buckets.  However, recall that we also need to
compute the distribution for the size of the result at each node.  This takes
time $O(\nbg{|A|} \nbg{|B|} \nbg{\sigma})$.
Can we do better?  Again, for specific distributions, 
it may be possible.
However, we can say more.  Suppose for uniformity we decide to have $b$ buckets
at every node for each variable.  If we have $b$ buckets for each of the
variables $|A|$, $|B|$, and $\sigma$, then we can get as many as $b^3$ buckets
for the size of the join $A \join B$.  To maintain $b$ buckets, we would have
to 
``rebucket''
after computing the probability.  Instead of 
rebucketing
after doing the computation, we can 
rebucket
each of $|A|$, $|B|$, and
$\sigma$ so that they have $\sqrt[3]{b}$ buckets.  Then the whole computation
takes time $O(b)$, as desired, and we have $b$ buckets for $|A \join B|$.  More
generally,
if we 
rebucket
each of $|A|$, $|B|$, and $\sigma$ so that they have
$\sqrt[3]{\nbg{|A|}}$, $\sqrt[3]{\nbg{|B|}}$, and $\sqrt[3]{\nbg{\sigma}}$
buckets, respectively, then we can carry out the computation in time
$O(\sqrt[3]{\nbg{|A|}\nbg{|B|}\nbg{\sigma}} + \nbg{|A|} + \nbg{|B|} +
\nbg{\sigma}) = O(\nbg{|A|} + \nbg{|B|} + \nbg{\sigma})$ steps.

\subsection{Strategies for Partitioning the Parameter Space}
\label{sec:tuning}
As we have seen, the complexity of all our 
algorithms
to computing or approximating the LEC plan depends on partitioning the
parameter space into a number of buckets.
A large number of buckets gives a closer approximation to the true probability
distribution, leading to a better estimate of the LEC plan.  On the other hand,
a smaller number of buckets makes the optimization process less expensive.
(As we mentioned earlier, the algorithm with one bucket reduces to the standard
System~R algorithm.)
For specific examples (such as Example~\ref{xam1}), choosing the buckets is
straightforward, as a function of the join algorithms being considered and the
sizes of the relations being joined.  We do not yet have a
general mechanism for choosing buckets optimally and efficiently.  However, we
have insights that should help us explore this issue in future work.

Consider our first two 
algorithms
(Sections~\ref{sec:lec:blackbox} and~\ref{sec:lec:cand}) which used the
System~R algorithm on a number of parameter settings to generate candidate
plans, and then evaluated each of these candidates to find the one of least
expected cost.  The partitioned parameter distributions are used in two ways:
the first is to generate candidate plans (by computing the best plan or $c$
best plans for each parameter value
considered)
and the second
is in computing the actual expected cost of the candidates generated.
Different, but related, issues arise in each case.  When generating
candidates, we are basically interested in determining the region of
parameter space in which to search for good candidates.  We can partition it
coarsely at first, and then generate more candidates in the region of
parameter
space that 
seems
most likely to contain the best candidates.  When computing
costs, recall that our goal is to find the candidate of least expected cost.
We do not always need an extremely accurate estimate of the cost to do this.
We expect to be able to associate a degree of accuracy with each
particular
partitioning---that 
is, a guarantee that the estimated expected cost of a plan using this
partitioning is within a certain degree of the true expected cost.  We may
be able to use  coarse bucketing to eliminate many plans and then use a
more refined bucketing to decide among the remaining few plans.

This insight applies even more to our third algorithm
(Section~\ref{generic}),
which 
actually computes the LEC plan.  If there are $j$ algorithms being compared at
a given node in the dag, the expected cost of only one of them (the algorithm
of least cost) needs to be computed accurately, since the other plans are
pruned. With respect to the pruned plans, we simply need to be satisfied that
their expected costs are higher than the chosen plan. Consequently, we can
start with a coarse bucketing strategy to do the pruning, and then refine the
buckets
as necessary.

Moreover, note that we do not have to use the same partitioning strategy at
every node.  We should use the partition appropriate to the strategies being
considered at that node.  The cost formulas of the common join algorithms are
very simple, at least with respect to a parameter like available memory.
As we saw, for fixed relation sizes, the cost for a sort-merge join 
has one of three possible values,
depending on the relationship between the memory and the size of the
larger relation; similarly, the cost of a nested-loop join has only one
of two possible values.  
Consequently, if we are considering a sort-merge join (resp.,
nested-loop join) for fixed relation sizes, we need deal with only three
(resp., two) buckets for memory sizes.
In general, we expect that we will be able to use features
of the cost formulas to reduce the number of buckets needed {\em on a
per-algorithm per-node basis}.

Another way to approach bucketing is to realize that, ultimately, we want
to compute the expected cost.  Fix a plan $\pl$.  Note that we can express
$\ec(\pl)$ as follows:
\[
\ec(\pl) = \sum_{c=0}^\infty c \Pr(\cst(\pl,\env) = c).
\]
For a fixed $c$, values of $\env$ that yields $\cst(\pl,\env) = c$ are called a
\emph{level set} of $\cst(\pl,\env)$.  If the cost of $\pl$ has relatively few
level sets, then it may be wise to bucket the parameter space with these level
sets in mind.  Suppose $\cst(\pl,\env)$ has $\ell$ level sets. In principle, we
can compute $\ec(\pl)$ with $\ell$ evaluations of the cost function, $\ell$
multiplications, and $\ell - 1$ additions.\footnote{This is also a lower bound,
if we want the exact expected cost (and if we treat the cost function as a
black box).}  We can do this if we have the probabilities for each level set.
In general the buckets will not correspond to level sets and we may evaluate
the cost function many times only to get the same answer each time.  So if we
bucket the joint distribution by using the level sets (instead of bucket each
parameter separately), we can minimize the computation involved in computing
the expected cost.  The cost function may have many level sets.  If we are
willing to settle for an approximate answer, we can bucket the range of the
function, thereby coalescing some of the level sets.  One problem with this
approach is that the probability of each level set may not be easy to
determine.







\section{Concluding Remarks}
\label{s:con}

This paper presents the simple idea of searching for query execution plans with
the least expected cost. To the best of our knowledge, this is a new approach
to query optimization, departing from the established approach used for the
past
two decades.  
Our approach can be viewed as a generalization of the traditional approach in the
sense that the traditional approach is essentially our approach restricted to
one bucket (in the static case).
Although this paper proposes the approach, we are aware that many
details need to be
worked out.
To make our task easier, we have made
simplifying assumptions. We now revisit some of these assumptions.
\begin{itemize}
\item Our presentation of the System~R query optimization algorithm is
rather
simplistic. The major issue we do not consider is parallelism, which can
play a
role in two ways (either through bushy join trees or through pipeline
parallelism).
In both cases, there is an interaction between the parallel components in terms
of memory used. While we have ignored this issue, current query optimizers do
model it, and we believe the same techniques can be applied to LEC optimization
as well.
\item
In dealing with changes during the execution of the plan, we made the
simplifying assumption that no change occurs during any one join ``phase''.
This is clearly an approximation of reality. Further, pipelined joins should be
treated together as a single phase while other algorithms (like a sort-merge
join) may involve multiple phases.
While we have certainly made simplifying assumptions here, we note that our
approach at least allows us to tackle a problem
not addressed
by other works in this area.
\item
When we considered multiple parameters, we assumed that the parameters were
independent.  This may not always be a reasonable assumption in practice.  It
would be of interest to see to what extent we could extend our techniques to
situations were there are some dependencies between the variables.
\end{itemize}
Although there is clearly work that needs to be done before we can use LEC
query optimization in production database systems, we believe that it is an
approach well worth exploring.  We are currently prototyping the algorithm of
Section~\ref{generic} to test its benefits against realistic queries and
execution environments. Such a prototype will also be useful to investigate the
impact of bucket choice (see Section~\ref{sec:tuning}) on the quality of LEC
plans.

\bibliographystyle{alpha}
\bibliography{z,joe,praveen}

\newcommand{\etalchar}[1]{$^{#1}$}
\begin{thebibliography}{SAC{\etalchar{+}}79}

\bibitem[Ant93]{rdb:dyn}
Gennady Antonshenkov.
\newblock {Dynamic Query Optimization in Rdb/VMS}.
\newblock In {\em Proceedings of the Ninth IEEE Conference on Data Engineering,
  Taipei, Taiwan}, pages 538--547, 1993.

\bibitem[GC94]{graefe:dynamic}
Goetz Graefe and Richard Cole.
\newblock {Optimization of Dynamic Query Evaluation Plans}.
\newblock In {\em {Proceedings of ACM SIGMOD '94 International Conference on
  Management of Data, Minneapolis, MN}}, 1994.

\bibitem[IK90]{Ioannidis90}
Y.E. Ioannidis and Y.C. Kang.
\newblock {Randomized Algorithms for Optimizing Large Join Queries}.
\newblock In {\em Proceedings of the ACM SIGMOD Conference on Management of
  Data}, pages 312--321, 1990.

\bibitem[Ill94]{illustra:usermanual}
Illustra Information Technologies, Inc, 1111 Broadway, Suite 2000, Oakland, CA
  94607.
\newblock {\em Illustra User's Guide}, June 1994.

\bibitem[INSS92]{Yannis:parametric}
Y.~Ioannidis, R.~Ng, K.~Shim, and T.K. Sellis.
\newblock {Parametric Query Optimization}.
\newblock In {\em Proceedings of the Eighteenth International Conference on
  Very Large Databases (VLDB), Vancouver, Canada}, pages 103--114, 1992.

\bibitem[KD98]{Kabra98}
Navin Kabra and David DeWitt.
\newblock {Efficient Mid-Query Re-Optimization of Sub-Optimal Query Plans}.
\newblock In {\em Proceedings of the ACM SIGMOD Conference on Management of
  Data}, pages 106--117, 1998.

\bibitem[LNSS93]{sampling:lnss}
Richard~J. Lipton, Jeffrey~F. Naughton, Donovan~A. Schneider, and S.~Seshadri.
\newblock {Efficient Sampling Strategies for Relational Database Operations}.
\newblock {\em {Theoretical Computer Science}}, pages 195--226, 1993.

\bibitem[Loh98]{lohman:pers98}
Guy Lohman, 1998.
\newblock Personal communication with Praveen Seshadri.

\bibitem[Pea88]{Pearl}
J.~Pearl.
\newblock {\em Probabilistic Reasoning in Intelligent Systems}.
\newblock Morgan Kaufmann, San Francisco, Calif., 1988.

\bibitem[PHH92]{Pirahesh:1}
Hamid Pirahesh, Joseph Hellerstein, and Waqar Hasan.
\newblock {Extensible/Rule Based Query Rewrite Optimization in Starburst}.
\newblock In {\em {Proceedings of ACM SIGMOD '92 International Conference on
  Management of Data, San Diego, CA}}, 1992.

\bibitem[PIHS96]{Poosala96}
V.~Poosala, Y.E. Ioannidis, P.J. Haas, and E.J. Shekita.
\newblock {Improved Histograms for Selectivity Estimation of Range Predicates}.
\newblock In {\em Proceedings of the ACM SIGMOD Conference on Management of
  Data}, pages 294--305, 1996.

\bibitem[Res87]{res}
M.~D. Resnik.
\newblock {\em Choices: An Introduction to Decision Theory}.
\newblock University of Minnesota Press, Minneapolis, 1987.

\bibitem[SAC{\etalchar{+}}79]{Selinger:1}
Patricia~G. Selinger, M.~Astrahan, D.~Chamberlin, Raymond Lorie, and T.~Price.
\newblock {Access Path Selection in a Relational Database Management System}.
\newblock In {\em {Proceedings of ACM SIGMOD '79 International Conference on
  Management of Data}}, pages 23--34, 1979.

\bibitem[SBM93]{sbm93}
K.~D. Seppi, J.~W. Barnes, and C.~N. Morris.
\newblock A bayesian approach to database query optimization.
\newblock {\em ORSA Journal on Computing}, 5(4):410--419, 1993.

\bibitem[Sha86]{shapiro:joins}
Leonard Shapiro.
\newblock {Join Processing in Database Systems with Large Main Memories}.
\newblock {\em {ACM Transactions on Database Systems}}, 11(3):239--264,
  September 1986.

\bibitem[Swa89]{swami:sigmod89}
Arun Swami.
\newblock {Optimization of Large Join Queries: Combining Heuristic and
  Combinatorial Techniques}.
\newblock In {\em Proceedings of the ACM SIGMOD Conference on Management of
  Data}, pages 367--376, 1989.

\bibitem[UFA98]{ufa98}
T.~Urhan, M.~J. Franklin, and L.~Amsaleg.
\newblock Cost based query scrambling for initial delays.
\newblock In {\em Proceedings of the ACM SIGMOD International Conference on
  Management of Data}, pages 130--141, Seatle, WA, June 1998.

\end{thebibliography}
\end{document}